# Locking ssDNA in a Graphene-Terraces Nanopore and Steering Its Step-by-Step Transportation via Electric Trigger

Wenping Lv *,a, Jiaxi Peng a, Dongsheng Xu b, Ren'an Wu*, a a CAS Key Laboratory of Separation Science for Analytical Chemistry, National Chromatographic R & A Center, Dalian Institute of Chemical Physics, Chinese Academy of Sciences (CAS), Dalian 116023, China

b Institute of Metal Research, Chinese Academy of Sciences (CAS), Shengyang110016, China

**ABSTRACT:** This study demonstrates that the nanopore terraces constructed on a multilayer graphene sheet could be employed to control the conformation and transportation of an ssDNA for nanopore sequencing. As adsorbed on a terraced graphene nanopore, the ssDNA has no in-plane swing nearby the nanopore, and can be locked on graphene terraces in a stretched conformation. Under biasing, the accumulated ions near the nanopore promote the translocation of the locked ssDNA, and also disturb the balance between the driven force and resistance force acted on the nucleotide in pore. A critical force is found to be necessary in trigging the kickoff of the ssDNA translocation, implying an inherent "field effect" of the terraced graphene nanopore. By changing the intensities of electric field as trigger signal, the "stop" and "go" of an ssDNA in the nanopore are manipulated at single nucleobase level. The velocity of ssDNA in the nanopore can also be regulated by the frequency of the electro-stimulations. As a result, a new scheme of controllable translocation of ssDNA in graphene nanopores is realized by introducing controllers and triggers, appealing more explorations in experiment.

Nanopore revolution is still in need in the sequencing of label-free DNA at single nucleobase resolution [1-3]. Nanopores tailored on atomic-scale thin two-dimensional (2-D) materials such as graphene promise the single-nucleobase recognition ability due to the comparable dimensions between the interval of adjacent nucleobases and the thickness of nanopore [4-6]. By using graphene nanopores, the intrinsic and induced disparities of local conformations of a DNA during translocation could be revealed in the fluctuations of ionic current and transversal conductance [4-7]. However, DNA sequencing has still not been achieved by a graphene nanopore due to the high sensitivity of these signals toward the stochastic fluctuations of DNA conformation in pore [5,7], and even the flexibility of graphene membrane itself [8]. To improve the resolution of nanopore sequencers further, it is first vital to overcome the fluctuations of DNA conformation in nanopores.

It is well known that the nucleobases in DNA could be adsorbed on hydrophobic $sp^2$-carbon surfaces [9], especially carbon nanotubes and graphene. Initially, experimentalists were focused on avoiding such an adherency of DNA on graphene because of the low capture ability of the nanopore towards an adsorbed DNA [10], but, the transportation of either dsDNA or ssDNA in a graphene nanopore was found slowing down by π-π stacking interaction between nucleobases and graphene [11-13]. In particular, a step-wise translocation of an adsorbed ssDNA could be observed in graphene nanopores [13]. By changing the polarity of surface charges of graphene, a stop-and-go motion of an adsorbed ssDNA in nanopore could be realized further [14], as the stacking conformation of nucleobases on graphene-like surfaces was regulated by the charge density of carbon atoms [14,15]. However, the in-plane motion of ssDNA on graphene surface was out of control in these schemes [13,14]. Stochastic swings of ssDNA on graphene surface might not only induce the conformational and temporal indeterminacies of each nucleobase in nanopore, but also disturb a steady distribution of ions those close to graphene surfaces under bias voltage, inducing potential uncertainties/noises of nanopore sequencing.

Inspired by the terrace-effect in the drilling of nanopore on multilayer graphene [16], in this contribution, we proposed a nanopore device which has several graphene terraces nearby the pore (Figure 1). These graphene terraces were designed to lock the local conformation of ssDNA (colored in white, Figure 1). Meantime, the multilayer graphene was expected to reduce the 1/f noise of nanopore [8]. By means of molecular dynamics (MD) simulations, we found that the sliding of a ssDNA (poly-(ATGC)4) nearby the pore was inhibited effectively for the adsorption of nucleobases on different graphene terraces. By analyzing the effective driven forces acted on the ssDNA, we found that a "gate voltage" was necessary to trigger the transportation of ssDNA in the nanopore. As a demo, we manipulated the "stop" and "go" statuses of the adsorbed ssDNA in the terraced graphene nanopore using a straightforward trigger signal—electric field pulses. By integrating the effects of graphene terraces and external stimulations, a controllable translocation of ssDNA through nanopore was realized at single nucleobase level.

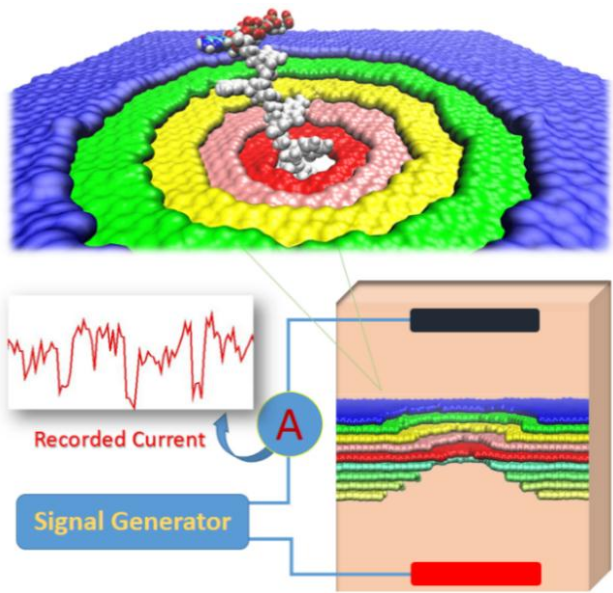

Figure 1. Schematic of a terraced graphene nanopore device. The highlighted drawing shows the locked portion (colored in white) of ssDNA on graphene terraces.

The stability of ssDNA (poly-(ATGC)4) in a terraced graphene nanopore was established by a comparison with the dynamics behavior of the ssDNA in a smooth graphene nanopore. Two nanopores were drilled in monolayer and multilayer graphene respectively, with the same aperture of 1.4 nm (Figures 2a-2b). Different with the smooth monolayer graphene nanopore (Figure 2a), four graphene terraces were tailored on two sides of the multilayer graphene membrane additionally (9-layers of graphene in total, Figure 1). The width of each terrace was about 0.7 nm (Supporting Information Figure S1), allowing the adsorption of single nucleobase (Figure 2b). Two 10-ns MD simulations were performed to monitor the conformational fluctuations of the ssDNA chain adsorbing on the two graphene nanopores in 1 M NaCl solution [11], respectively. In the simulations, only the carbon atoms at the edge of graphene were fixed [17], allowing the flexibility of the two graphene nanopores [8].

The sampled structures of ssDNA in the two graphene nanopores were shown as density maps in Figures 2c-2d. In line with a previous investigation [13], the free swing of nucleobases on the smooth graphene surface induced a broad range of ssDNA distribution (Figure 2c). In contrast, the terraces on multilayer graphene locked the local conformation of ssDNA on the surface (Figure 2d), inhibiting the fluctuation of nucleobases nearby the pore effectively (indexes #7-#13, Figure 2g). Besides, the backbone of this portion of locked ssDNA was also stretched relative to the swing ssDNA (Figure 2h). Such a "static" and "extended" conformation of an ssDNA on graphene terraces means a limited perturbation of the ssDNA towards ions near the pore. As expected, a high-density region of ions was observed inside the terraced nanopore for both $Na^+$ and $Cl^-$, but no observable concentration of either $Na^+$ or $Cl^-$ occurred inside the pore on monolayer graphene (Figures 2e-2f). These results indicate that the structural stabilities of ssDNA and ions could be both enhanced in the terraced graphene nanopore, promising in the improvement of the accuracy of sequencing by either ionic current or transversal tunneling conductance measurements [5,13,18].

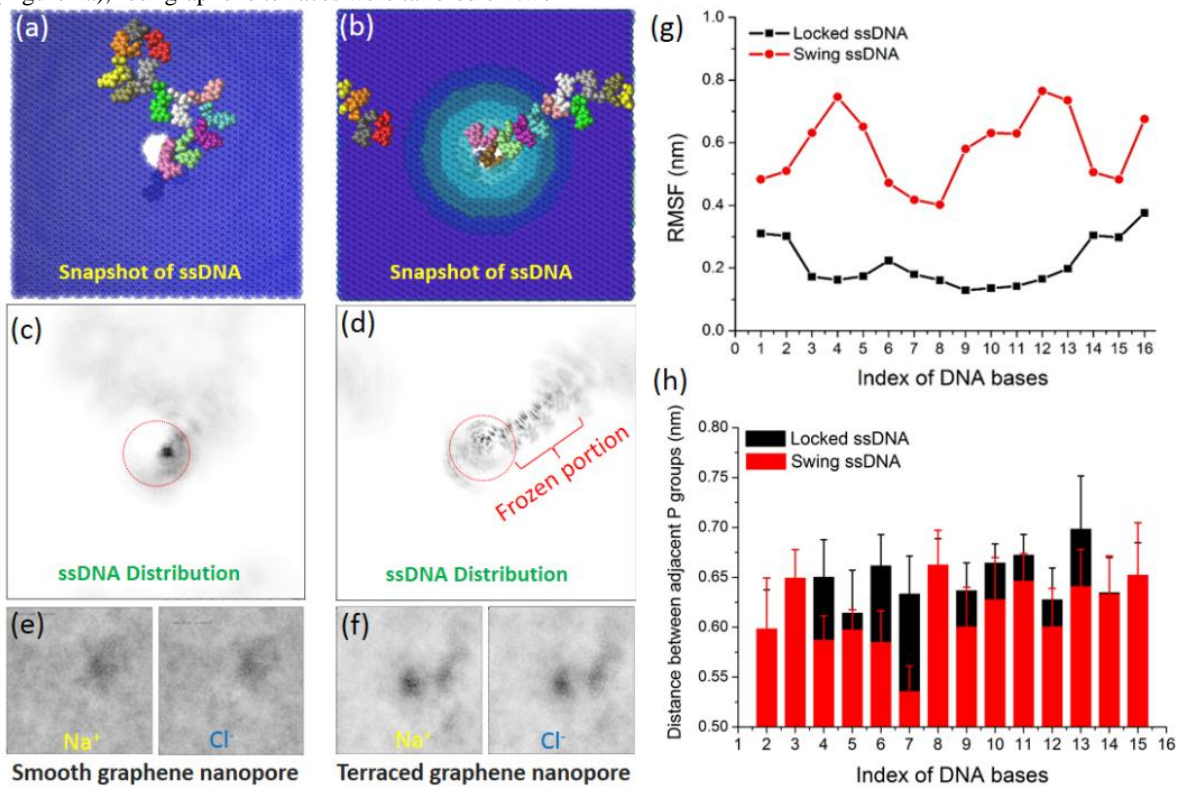

Figure 2. Snapshots of ssDNA adsorbing on a smooth graphene nanopore (a) and a terraced graphene nanopore (b), respectively. Density maps of ssDNA adsorbing on (c-d) and ions projected in (e-f) surfaces of the smooth and terraced graphene nanopore, respectively. (g) Root-mean-square fluctuations (RMSFs) of each nucleobase on the locked and swing ssDNA. (h) Average distance between two adjacent phosphate (**P**) groups on the backbone of locked and swing ssDNA.

The translocation of an adsorbed ssDNA through a graphene nanopore is dominated by the external driven force as well as the encountered resistive forces in system. In nanopore experiments, a bias voltage induced a driven electric field pulling the ssDNA through a nanopore. Due to the insulativity of graphene membrane in normal direction, the drop of voltage in nanopore experiments occurred on the immediate vicinity of the pore, and therefore electrical forces acted on a short part of ssDNA locally [19]. Herein, we assumed that the driven force applied externally ($F_d$), the influence of ions ($F_o$), and the resistive force caused by DNA-graphene interaction ($F_g$) were all acted on the nucleotide of ssDNA in pore (Figure 3a). Under a biasing electric field, the external driven force and the influence of ions could be considered together as an electrophoretic force $F_e=F_d+F_o$ ($F_o$ and $F_d$ were assumed in the same direction, and the effect of water was ignored). To trigger the ssDNA translocation in pore, $F_e \geq F_g$ should be achieved. At the critical (starting) state before a step-wise translocation of the ssDNA, the electrophoretic force $F_e$ is balanced with the resistive force $F_g$. Ideally, a nucleobase at the critical state in pore could alter in the statuses of desorption from and re-adsorption on the edge of graphene nanopore. To arrest this critical state of ssDNA translocation, a quasi-desorption state of the nucleobase from the edge of the pore was built by applying an adaptive force $F_a$ on the phosphate group of the nucleotide (Figure 3b), resulting in a new balance: $\lim_{desp} F_g = F_e + F_a$.

The measured $F_a$, $F_e$ and $F_o$ were plotted as the function of biasing electric field in Figure 3c (see **Computational details**). As these forces were balanced, when $F_e$ was increased with the enhancement of external electric field, $F_a$ was decreased correspondingly. The measured values of $F_a$ were fitted to a linear equation, with the weighting of the reversal of its fluctuations, showing that $F_a$ arrived to zero when the external electric field was about 1.1 V/nm. It corresponds to an effective driven force ($F_e$) around 400 pN. Below the critical point, the transportation of the adsorbed nucleobase in pore could not be kick-started. On the other hand, the fluctuations of $F_a$ under biasing were huge as compared with the fluctuations of $F_a$ without the external electric field. The huge fluctuations of $F_a$ around the critical point mean that the balance between $F_e$ and $F_g$ was easy to be disturbed under an external electric field.

Besides the external electric field, the accumulation of ions on graphene surface resulted in an additional voltage drop in the nanopore [17]. By filtering out the external driven force, which was exactly in line with the strength of applied electric field, the contribution of the ions towards ssDNA translocation were obtained (Figure 3c). We found that $F_o$ was positive under all applied external electric fields. In particular, when the external electric field was greater than 0.7 V/nm, $F_o$ even became stronger than the external driven force. Such a powerful contribution of the ions towards ssDNA translocation might be one origination of the fragile balance (fluctuations of $F_a$) of ssDNA translocation under biasing, because of the electrostatic turbulent fluctuations induced by the inherent mobility of the ions nearby the pore. Actually, the intermolecular interactions at the interface of graphene nanopore might be more complex due to the ignored factors in our assumption, *e.g.* the dipolar rearrangements of water molecules under the synergetic impacts from the hydrophobicity of graphene, external electric field, and electrostatic perturbation of ions and ssDNA.

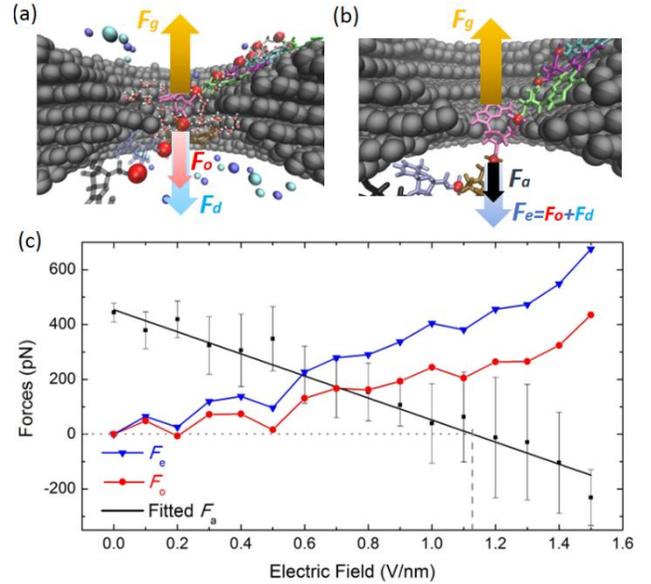

Figure 3. (a) A graphical representation of the driven force ($F_d$) and the encountered forces ($F_g$ and $F_o$) on the nucleotide of ssDNA in the terraced graphene nanopore. Here, we assumed that the $F_o$ was in the direction of $F_d$. The **P** atoms on ssDNA were highlighted by using red beads. (b) A nucleobase at the desorption status from the upper surface of the graphene nanopore. At this critical state, $\lim_{desp} F_g = F_e + F_a$ should be achieved. (c) Evolutions of $F_a$, $F_e$ and $F_o$ along the applied external biasing electric field.

The availability of the predicted critical electric field was assessed by performing an additional simulation for the same ssDNA in the terraced graphene nanopore under the external electric field of 1 V/nm (slightly smaller than 1.1 V/nm). As expected, no ssDNA translocation was observed in a 200-ns MD simulation (Supporting Information Figure S2). Different with the inconstant step-wise translocations of ssDNA in a smooth graphene nanopore [13], the employed graphene terraces can effectively restrain the uncertain translocations of ssDNA even though the applied electric field here was much higher relative to the previous study. Such a translocation character of a locked ssDNA in nanopore shows an inherent "field effect" of the terraced graphene nanopore, indicating that a "gate voltage" should be supplied to trigger a translocation-step of ssDNA. Also, it means that the "stop" and "go" statuses of an adsorbed ssDNA in the terraced graphene nanopore might be controlled by introducing a stimulation in the biasing electric field.

Therefore, a weak biasing electric field with temporal high strength was generated to steer the motion of the adsorbed ssDNA in the terraced graphene nanopore (Figure 4a). Baseline of the external electric field ($E_b$) was set as 0.2 V/nm, which was commonly used in ionic current measurements [5,13]. Each stimulation has a strength of 5 V/nm ($E_p$) and a width of 2 ps ($T_p$). Four translocation simulations were carried out with different time interval ($T_b$) between two impulses from 100 ps, 200 ps and 500 ps, to 1 ns, respectively. We found that a step-by-step translocation of ssDNA in pore is occurred according to the frequency of stimulations (Figure 4b). That is, the nucleobase in pore has no significant shift in the direction of biasing between two impulses (highlighted in red dot boxes, Figure 4b), and all translocations of the nucleobase were finished within the time duration of a stimulation (Supporting Information Movies S1-S2).

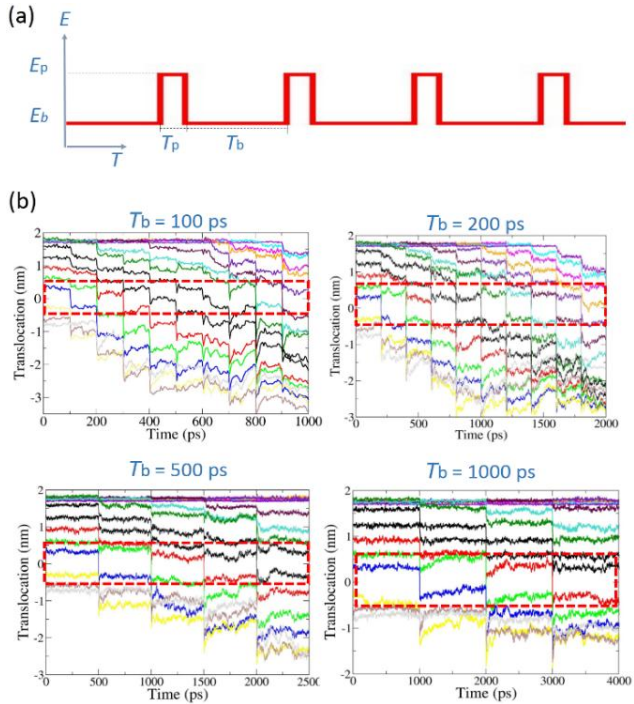

Figure 4. (a) A graphic presentation of the electric filed pulses. (b) Time evolutions of the translocation of each nucleobase under the pulsed electric field with $T_p$ of 100 ps, 200 ps, 500 ps, and 1000 ps, respectively. The entrance of the terraced graphene nanopore was marked with red dot boxes.

The quick motions of one nucleobase in pore before and during an electric stimulation were presented in Figure 5. We found that the nucleobase underwent a process of slipping off from the edge of graphene and entering the nanopore during a stimulation. Interestingly, due to the elasticity of the ssDNA backbone, a hysteretic change of conformation was occurred on the nucleobase. For instance, in the first 0.8 ps of a stimulation, no significant movement of the nucleobase was observed, because the electric force was mainly acted on the charged phosphate group of ssDNA backbone. In addition, an obvious conformational warp of the nucleobase was occurred at 1.2 ps of stimulation, and the nucleobase was altered in the adsorption and desorption statuses on graphene edge during an additional stimulation of 0.4 ps. Finally, the nucleobase shifted into the pore successfully after the stimulation of 2 ps. These results indicate that the "stop" and "go" of an adsorbed ssDNA could be manipulated at single nucleobase level by employing an electric stimulation. Thus, the overall transportation speed of an ssDNA in pore might also be regulated by the frequency of stimulations exactly.

To assess the efficiency of these stimulations, 50 times of repetition simulation were performed further, with a $T_b$ of 100 ps for saving computational time. We found that 383 translocation events were triggered successfully after 450 stimulations, demonstrating a driven efficiency over 85%. The details of all these translocations under each pulse were tabulated in Figure 6a, showing that the range of nucleobases triggered by each pulse was kept steady until the sixth pulse. Correspondingly, the number of skips of nucleobases (colored in green) after each pulse were conserved in range of 3 to 14 until the sixth pulse, maintaining the stability of the amount of the expected translocations (highlighted in red dot box). The wider range of the triggered nucleobases in the last four pulses (6~9) might be a result of the weakened restriction of the graphene terraces on the nucleobases after portion of the ssDNA was transported through the nanopore (Supporting Information Movie S2).

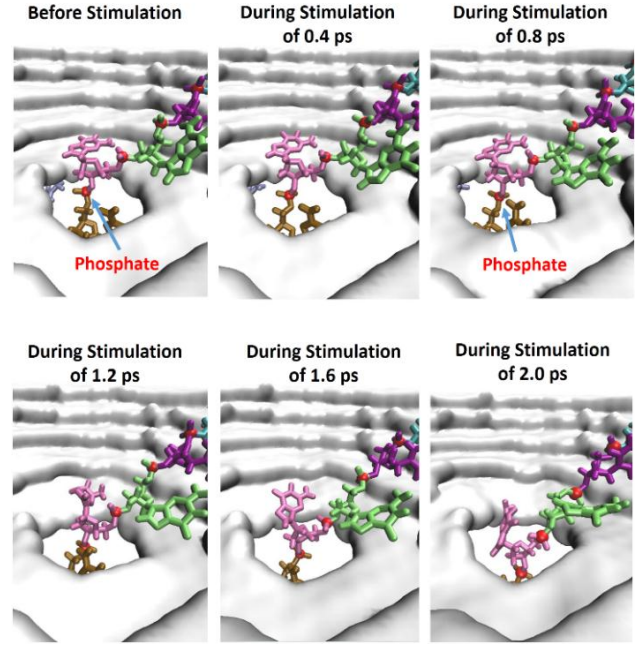

Figure 5. A typical process of the conformational transformation of a nucleobase at the entrance of nanopore before and during an electric stimulation.

As a comparison, the identical electric stimulations were generated to steer the transportation of the same ssDNA in a smooth graphene nanopore which has the same aperture. Results demonstrated that 50 times of stimulation triggered only 35 times of expected translocation for the first nucleobase in pore (R13) (Figure 6b). And, a rapid decline of the number of expected translocations and wider skip ranges of the nucleobases were observed with the increase of the number of pulses. The wide range of skips means that the triggered translocation of nucleobases in a smooth graphene nanopore was out of control comparing with that in the terraced graphene nanopore. Actually, one stimulation could activate several nucleobases nearby the smooth graphene nanopore (inset of Figure 6b), and the adjacent nucleobases would transform into a stacking conformation nearby the smooth nanopore (Supporting Information Movie S3). These observations revealed a drawback of the conformational variations of ssDNA induced by its free sliding nearby the nanopore. Therefore, to steer the "stop" and "go" of an ssDNA chain in a graphene nanopore at single nucleobase level, the locking of nucleobases by graphene terraces is as important as the applying of external stimulations.

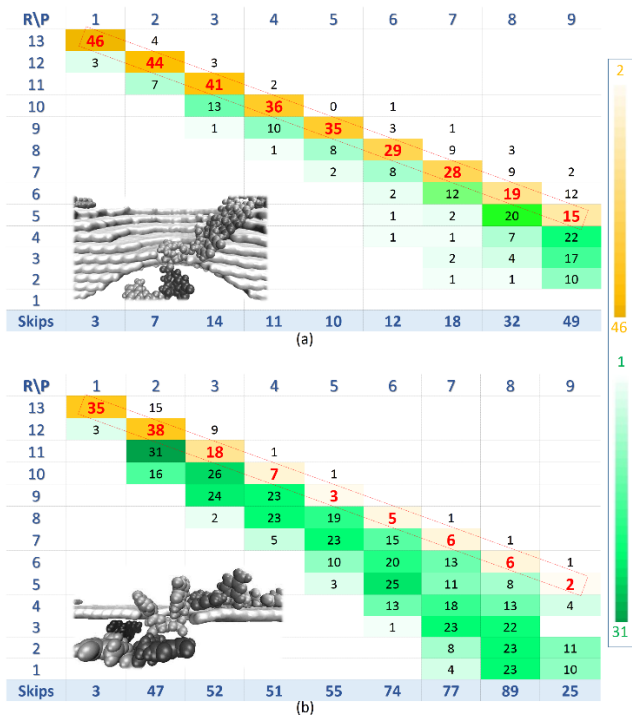

Figure 6. The statistics of triggered translocation events of each nucleobase (from 13 to 1) induced by each electric pulse (from 1 to 9) in the 50 repeating simulations in the terraced (a) and smooth (b) graphene nanopores, respectively. The expected translocations of nucleobases under each pulse were highlighted in red dot boxes. The skipped translocations of nucleobases under each pulse were colored in green.

Furthermore, we explored the details of the failed stimulations, by examining the trajectory and structure evolutions of nucleobases after one electric pulse. Position fluctuations of the nucleobase in pore along the direction of electric field were presented in Figure 7a. In the 50 times of simulation, 4 inefficient stimulations were occurred (the barycenter of nucleobase was on the upper surface of graphene nanopore during the $T_b$ of 100 ps, Figure 7a). Interestingly, we found that there was no significant conformational difference of ssDNA between a successful and the 4 failed stimulations before the triggering (Supporting Information Figure S3). It means that the inefficient stimulations might be stem from the conformational differences of the ssDNA during $T_p$. Actually, the nucleobase in pore underwent different evolution processes during the these stimulations (Supporting Information Movies S4-S8). The nucleobase under the 4 failed stimulations could not get rid of the interaction of graphene after $T_p$ (Figure 7b), i.e, the nucleobase was sticking on the rim of graphene nanopore during the $T_b$ of 100 ps (31# trajectory). It implying that the edge friction between nucleobase and graphene also restricted the transportation of ssDNA. For other 3 stimulations (32#, 0# and 10# trajectories), two nucleobases were shifted into the nanopore simultaneously, resulting a jam of transportation. Especially, the closely contact between two nucleobases would induce a stacking conformation (0# and 10# trajectories). Fortunately, these transportation accidents were not frequent for the locked ssDNA in the terraced graphene nanopore (only 8%). A careful optimization of the structure of graphene terraces and the stimulation signals should be beneficial to further inhibit these unexpected transportation accidents.

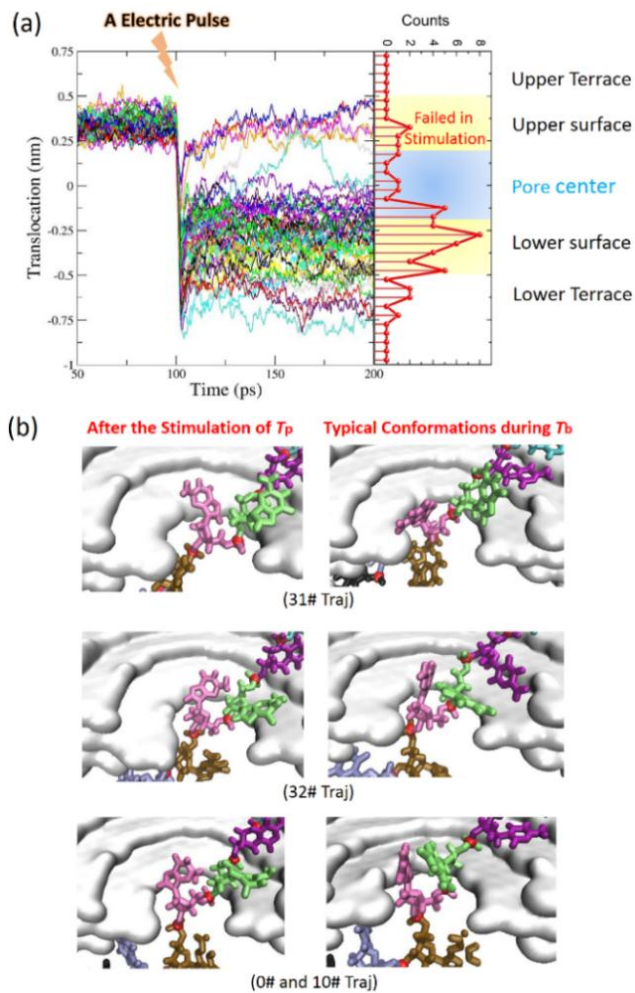

Figure 7. (a) The translocations of a nucleobase before and after an electric stimulation (50 times of repetition). (b) The final conformations of the nucleobase after the four failed stimulations of $T_p$, and the typical conformations of the nucleobase during $T_b$ of 100 ps.

In practice, the graphene terraces on such a nanopore could be constructed and optimized by using the single-atom resolution tailoring technologies [20,21], or by a programed *in-situ* synthesis strategy [22], and even by an assembly of individual graphene nanopore structures layer-by-layer [23]. Besides, chemical functionalization and hydrophobicity tailoring [24] could be applied on these graphene terraces, to enhance the in-plane friction and/or to improve the discrimination capability of a nanopore to different kinds of nucleobases. To push the ssDNA away from the locked configuration using a low frequency (below kHz) stimulation, a long enough sensing time could be supplied to recognize each nucleobase in nanopore [13]. The integration of multiple sensing strategies, such as ionic current and transversal tunneling current, is also a route to improve the error-correction ability of a single translocation event [4,7,18,25,26]. New technologies should be developed to measure the conductance of each layer of graphene nanopore with high quality, and an inserting of interlayer material between two graphene terraces might be helpful on this point [12,27].

In summary, the proposed structure of a terraced graphene nanopore could be used to manipulate the transportation of an adsorbed ssDNA at single nucleobase level. The structure of ssDNA was locked by graphene terraces at the "stop" state, efficiently avoided the random swings of ssDNA nearby the nanopore. Such a stable adsorption of ssDNA on graphene terraces also provided a basis for reliable sequencing by transversal conduction measurements. A "field effect" of the terraced graphene nanopore in ssDNA translocation was existed because a critical force was necessary to kick-start the translocation ("go") of the adsorbed nucleobases on ssDNA. By introducing external stimulations, the step-by-step transportation of the locked ssDNA in the terraced graphene nanopore could be manipulated effectively at single nucleobase level. Such a steerable transportation of a locked ssDNA through nanopore simplified the temporal-spatial complexity of ssDNA translocation, should be beneficial to improve the reliability of nanopore sequencing.

**Computational Details.** In the calculation of effective forces, $\lim_{desp} F_g$ was considered as a constant for a certain conformation of ssDNA adsorbing on graphene. $F_e$ was obtained by measuring $F_a$. In our simulations, the external driven force was in proportion to the applied biasing electric field ($F_d = qE_z$, $q$ is the net charge on one phosphate group), therefore, $F_o$ was obtained ($F_o = F_e - F_d$). Measurement of adaptive force $F_a$ was carried out by using a spring with a constant of 830 pN/nm. The critical state of ssDNA in nanopore was defined as that the relative displacement of the nucleobase and the graphene in $z$-direction was in range of 0.45 nm - 0.55 nm. Correspondingly, $F_a$ was measured as the average force in this range. All MD simulations were performed using GMXMACS 4.5 [28] with AMBER force field [29] and TIP3P water model [30]. Simulation box was about $10 \times 10 \times 8$ nm$^3$, and 3-D periodic boundary condition (PBC) was employed in simulations. Trajectories were collected in canonical ensembles at 300 K via a V-rescale [31] heating bath. Detailed simulation procedures and other MD parameters were same as previous literatures [5,11].

## ASSOCIATED CONTENT

**Supporting Information**. The atomic structure of the two graphene nanopores, the translocation behavior of ssDNA under the biasing of 1 v/nm, the conformational difference of ssDNA between a successful and the four failed stimulations before the triggering, and the movies of ssDNA translocation in the two nanopores.

## AUTHOR INFORMATION

**Corresponding Authors**

* W.P.L. Email: wenping@dicp.ac.cn

* R.A.W. Email: wurenan@dicp.ac.cn

**Notes**
The authors declare no competing financial interest.

## ACKNOWLEDGMENT
We deeply thank Prof. Dr. Klaus Schulten, his postdocs, and Dr. Alek Aksimentiev from University of Illinois at Urbana-Champaign, and Dr. Meni Wanunu from Northeastern University for their critical reading and comments on the manuscript. We thank Miss. Lihua Ye for the convenience of data transportation. This work was supported by the financial support from the National Natural Science Foundation of China (Nos. 21175134 and 21375125) and the Creative Research Group Project of National Natural Science Foundation of China (No. 21321064).

## REFERENCES

1   Wanunu, M. Nanopores: A journey towards DNA sequencing. *Phys Life Rev* **9**, 125-158, (2012).
2   Spencer, C. & Meni, W. Challenges in DNA motion control and sequence readout using nanopore devices. *Nanotechnology* **26**, 074004 (2015).
3   Pennisi, E. GENOMICS DNA Sequencers Still Waiting for The Nanopore Revolution. *Science* **343**, 829-830 (2014).
4   Traversi, F. *et al.* Detecting the translocation of DNA through a nanopore using graphene nanoribbons. *Nat Nanotechnol* **8**, 939-945, (2013).
5   Lv, W. P., Liu, S. J., Li, X. & Wu, R. A. Spatial blockage of ionic current for electrophoretic translocation of DNA through a graphene nanopore. *Electrophoresis* **35**, 1144-1151, (2014).
6   Garaj, S. *et al.* Graphene as a subnanometre transelectrode membrane. *Nature* **467**, 190-U173, (2010).
7   Girdhar, A., Sathe, C., Schulten, K. & Leburton, J. P. Graphene quantum point contact transistor for DNA sensing. *P Natl Acad Sci USA* **110**, 16748-16753, (2013).
8   Heerema, S. J. *et al.* 1/f noise in graphene nanopores. *Nanotechnology* **26**, 074001 (2015).
9   Zheng, M. *et al.* DNA-assisted dispersion and separation of carbon nanotubes. *Nat Mater* **2**, 338-342, (2003).
10  Merchant, C. A. *et al.* DNA Translocation through Graphene Nanopores. *Nano Lett* **10**, 2915-2921, (2010).
11  Lv, W. P., Chen, M. D. & Wu, R. A. The impact of the number of layers of a graphene nanopore on DNA translocation. *Soft Matter* **9**, 960-966, (2013).
12  Banerjee, S. *et al.* Slowing DNA Transport Using Graphene–DNA Interactions. *Advanced Functional Materials*, 201403719 (2014).
13  Wells, D. B., Belkin, M., Comer, J. & Aksimentiev, A. Assessing Graphene Nanopores for Sequencing DNA. *Nano Lett* **12**, 4117-4123, 301655d (2012).
14  Shankla, M. & Aksimentiev, A. Conformational transitions and stop-and-go nanopore transport of single-stranded DNA on charged graphene. *Nat Commun* **5**, Artn 5171 (2014).
15  Lv, W. P. The adsorption of DNA bases on neutral and charged (8,8) carbon-nanotubes. *Chem Phys Lett* **514**, 311-316, (2011).
16  Fischbein, M. D. & Drndic, M. Electron beam nanosculpting of suspended graphene sheets. *Appl Phys Lett* **93**, Artn 113107 (2008).

17  Sathe, C., Zou, X. Q., Leburton, J. P. & Schulten, K. Computational Investigation of DNA Detection Using Graphene Nanopores. *Acs Nano* **5**, 8842-8851, (2011).
18  Chaitanya, S., Anuj, G., Jean-Pierre, L. & Klaus, S. Electronic detection of dsDNA transition from helical to zipper conformation using graphene nanopores. *Nanotechnology* **25**, 445105 (2014).
19  van Dorp, S., Keyser, U. F., Dekker, N. H., Dekker, C. & Lemay, S. G. Origin of the electrophoretic force on DNA in solid-state nanopores. *Nat Phys* **5**, 347-351, (2009).
20  Russo, C. J. & Golovchenko, J. A. Atom-by-atom nucleation and growth of graphene nanopores. *P Natl Acad Sci USA* **109**, 5953-5957, (2012).
21  Yin, R. C. *et al.* Ascorbic Acid Enhances Tet-Mediated 5-Methylcytosine Oxidation and Promotes DNA Demethylation in Mammals. *J Am Chem Soc* **135**, 10396-10403, (2013).
22  Waduge, P., Larkin, J., Upmanyu, M., Kar, S. & Wanunu, M. Programmed synthesis of freestanding graphene nanomembrane arrays. *Small* **11**, 597-603, (2015).
23  Lv, W. & Wu, R. a. The interfacial-organized monolayer water film (MWF) induced "two-step" aggregation of nanographene: both in stacking and sliding assembly pathways. *Nanoscale* **5**, 2765-2775, (2013).
24  Schneider, G. F. *et al.* Tailoring the hydrophobicity of graphene for its use as nanopores for DNA translocation. *Nat Commun* **4**, Artn 2619, (2013).
25  Avdoshenko, S. M. *et al.* Dynamic and Electronic Transport Properties of DNA Translocation through Graphene Nanopores. *Nano Lett* **13**, 1969-1976, (2013).
26  Qiu, W. Z. & Skafidas, E. Detection of Protein Conformational Changes with Multilayer Graphene Nanopore Sensors. *Acs Appl Mater Inter* **6**, 16777-16781, (2014).
27  Venkatesan, B. M. *et al.* Stacked Graphene-Al2O3 Nanopore Sensors for Sensitive Detection of DNA and DNA-Protein Complexes. *Acs Nano* **6**, 441-450, (2012).
28  Hess, B., Kutzner, C., van der Spoel, D. & Lindahl, E. GROMACS 4: Algorithms for Highly Efficient, Load-Balanced, and Scalable Molecular Simulation. *Journal of Chemical Theory and Computation* **4**, 435-447, (2008).
29  Cornell, W. D. *et al.* A 2nd Generation Force-Field for the Simulation of Proteins, Nucleic-Acids, and Organic-Molecules. *J Am Chem Soc* **117**, 5179-5197 (1995).
30  Jorgensen, W. L., Chandrasekhar, J., Madura, J. D., Impey, R. W. & Klein, M. L. Comparison of Simple Potential Functions for Simulating Liquid Water. *J Chem Phys* **79**, 926-935, (1983).
31  Bussi, G., Donadio, D. & Parrinello, M. Canonical sampling through velocity rescaling. *J Chem Phys* **126**, Artn 014101, (2007).